\documentclass{JHEP3}
\usepackage{graphics}
\title{Semi-Classical Mechanics in Phase Space: \\ The Quantum Target of Minimal Strings } 
\author{C\'esar G\'omez, Sergio Monta\~nez and Pedro Resco  \\ {\it Instituto de F\'\i 
sica Te\'orica CSIC/UAM,}\\ {\it C-XVI Universidad Aut\'onoma,}\\ {\it E-28049 Madrid \rm 
SPAIN}\\
E-mail: \email{cesar.gomez@uam.es}, \email{sergio.montannez@uam.es},   
\email{juanpedro.resco@uam.es} }

\abstract{The target space $M_{p,q}$ of $(p,q)$ minimal strings is embedded into the phase 
space of an associated integrable classical mechanical model. This map is derived from the 
matrix model representation of minimal strings. Quantum effects on the target space are 
obtained from the semiclassical mechanics in phase space as described by the Wigner 
function. In the classical limit the target space is a fold catastrophe of the Wigner 
function that is smoothed out by quantum effects. Double scaling limit is obtained by 
resolving the singularity of the Wigner function. The quantization rules for backgrounds 
with ZZ branes are also derived.}

\preprint{
IFT 05/30\\
{\tt hep-th/0506159}
}

\keywords{String Theory}

\newcommand{\tr}{\ensuremath{\mathrm{tr}}}

\def\ben{\begin{equation}}
\def\een{\end{equation}}
\def\bea{\begin{eqnarray}}
\def\eea{\end{eqnarray}}

\makeindex
\begin{document}

\section{Introduction}

The recent progress in the study of Liouville theory \cite{Dorn:1994sv}
 opens the possibility to address some problems of quantum gravity   using as a 
theoretical laboratory minimal $(p,q)$ strings \cite{Seiberg:2003nm,McGreevy:2003kb}. The   
information about the minimal string target space is encoded in the   dynamics of the FZZT 
branes \cite{Maldacena:2004sn}. In this paper we   present a new approach to study quantum 
corrections to the minimal string target geometry for $(2,q)$ models.

The strategy we follow is to map the minimal string data about FZZT   brane amplitudes 
into an integrable classical mechanical model. By this   procedure we map the minimal 
string target into a curve in   the mechanical model phase space. The form of this map can 
be directly derived   from the matrix model representation of the minimal string. Once we 
have embedded the classical string target into the phase space of the auxiliary mechanical 
model, quantum effects on the target, i.e quantum gravity effects, are derived from the 
semi-classical mechanics on phase space as it is described by the Wigner function of the 
mechanical system\footnote{The information encoded in the Wigner function is, in this 
interpretation, the same type of information you will expect to derive from a 
Hartle-Hawking wave function \cite{Ooguri:2004zv}.}. In the classical $h=0$ limit the 
support of the Wigner function is concentrated on the classical target submanifold. 
Moreover the classical target defines a catastrophe singularity for the Wigner function 
that is smoothed out by quantum effects \cite{Berry}. The semi-classical Wigner function 
spreads out on phase space in the way dictated by the type of singularity associated with 
the classical target. In particular, the exact double scaling limit is obtained by 
resolving the fold singularity of the Wigner function at the classical turning point. This 
leads to an interesting connection between catastrophe theory and double scaling limit. 
The Stokes's phenomenon described in \cite{Maldacena:2004sn} enters into this picture in a 
very natural way dictating the type of saddles contributing to the Wigner function in the 
classically forbidden region of phase space.

In this approach ZZ branes are related to the type of singularity at   the turning point. 
For backgrounds with ZZ branes a quantization   rule involving the number of ZZ branes 
emerges naturally. This leads   to fix the relation between the Planck constant of the 
auxiliary   mechanical model and the string coupling constant.

The plan of the paper is the following. In section 3 we introduce   the auxiliary 
mechanical model. In section 4 we define the Wigner   function and describe the type of 
singularity of this function on   target space. In section 5 we derive the map from 
minimal strings   into the auxiliary classical mechanical model from the matrix model   
description of minimal strings. In section 6 we derive the double   scaling limit solution 
by an uniform approximation to the Wigner   function near the singularity. We work out the 
gaussian model in   full detail. In section 7 we consider the issue of ZZ brane   
backgrounds, we derive the quantization rules and relate the type of   singularity at the 
classical turning point with the types of   different ZZ branes.

\section{The Classical Target of Minimal Strings}

For minimal strings defined by coupling $(p,q)$ minimal
models to Liouville theory, a Riemann surface $M_{p,q}$, playing the   role of a classical 
target space time, can be naturally defined   using the FZZT amplitude on the disk 
\cite{Seiberg:2003nm}. Let $\Phi(\mu, \mu_{b})$ be the FZZT disk amplitude for   $\mu$ and 
$\mu_{b}$ the bulk and boundary cosmological constants.   Defining \ben x= 
\frac{\mu_{b}}{\sqrt{\mu}} \een and \ben y=   \frac{1}{(\sqrt{\mu})^{1+\frac{1}{b^{2}}}} 
\partial_{x}\Phi \een the   Riemann surface $M_{p,q}$ is given by \ben
F(x,y) = T_{q}(x) - T_{p}(y) = 0
\een
where $T_{p}$ are the Chebyshev polynomials of the first kind. Singular points of 
$M_{p,q}$, defined by $F(x,y)= \partial_{x}F =   \partial_{y}F = 0$, are one to one 
related to eigenstates of the   ground ring (principal ZZ brane states). In this 
geometrical   interpretation we have \ben\label{uno}
\Phi(x) = (\sqrt{\mu})^{1+\frac{1}{b^{2}}} \int_{P}^{x}ydx
\een
and for the ZZ amplitude
\ben
\Phi_{ZZ} = (\sqrt{\mu})^{1+\frac{1}{b^{2}}} \oint ydx
\een
where the integration closed path passes through the singular point. As it is clear from 
(\ref{uno}), the curve $M_{p,q}$ can be   interpreted as the moduli of FZZT branes 
\cite{Maldacena:2004sn} and therefore as a model   for the target space of the 
corresponding minimal string.

\section{The Auxiliary Classical Mechanical System}

Let us consider the simplest integrable system with just one   constant of motion, the 
energy $E$, corresponding to a non   relativistic particle in a potential $V(q)$. Let us 
denote $p(q,E)$   the two valued function giving the momentum. We define the reduced   
action $S(q)$ as \ben \label{action} S(q, E) = \int_{P}^{q} p(q',E)dq' \een The one 
dimensional ``torus'' in phase space, which we will denote   $\Sigma_{E}$, is defined by 
the curve \ben p = p(q,E) \een The motion in phase space is confined to lay on this torus. 
For a   bounded system the action variable is defined by \ben I= \frac{1}{2\pi} \oint 
p(q)dq \een where the integral is on the homology cycles of the ``torus'',   i.e 
$\Sigma_{E}$ itself for the one dimensional case, and the angle   variable by \ben \theta 
=\partial_{I} S(q,I) \een For an unbounded system we define $I$ to be simply the energy 
$E$   and $\theta$ the uniformizing parameter of the open "torus".

In what follows we will associate with the $(2,q=2k-1)$ minimal string   data $\Phi(x)$ 
and the curve $M_{p,q}$ a classical one dimensional   unbounded integrable model by the 
following map \ben \label{dos}
\Phi(x) = i S(q,E=0)
\een
with $q=x$ and $S$ the analytic continuation of (\ref{action}). By   this map the curve 
$M_{p,q}$, defined by $y=y(x)$, is transformed into the curve  $p=p(q,E=0)$ in phase 
space, i.e the torus   associated with the integral of motion $E$.

\begin{figure}
\centering
\includegraphics{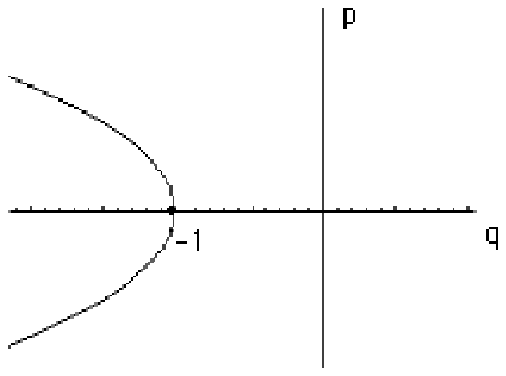}
\begin{center}
Figure~1: \footnotesize{Phase space curve $p(q,E=0)$ for (2,1) minimal string}
\end{center}
\end{figure}

As a concrete example, we can consider the curve $M_{2,1}$. It is   given by \ben y= \pm 
\sqrt{\frac{x+1}{2}} \een which leads, by using (\ref{dos}), to a classical mechanical 
model   defined by \ben \left[ V(q)-E \right] \arrowvert_{E=0}= -\frac{x}{2} - \frac{1}{2} 
\een with the curve $p(q,E=0)$ as depicted in figure 1. In general, for   the $(2,2k-1)$ 
minimal string we have \ben \label{tres} y= \pm 
\sqrt{2^{2k-3}(x+1)\prod_{n=1}^{k-1}(x-x_{1,n})^2}
\een
where $x_{1,l}=-\cos \frac{2\pi l}{2k-1}$. The curve $p(q,E=0)$ on   phase space has the 
structure depicted in figure 2. (\ref{tres}) has   singular points at $y=0$, $x=x_{1,l}$, 
denoting the existence of ZZ   brane states. The existence of these singular points is 
reflected in   the $p(q,E=0)$ curve as the extra points $A_l$ in figure 2.

\begin{figure}
\centering
\includegraphics{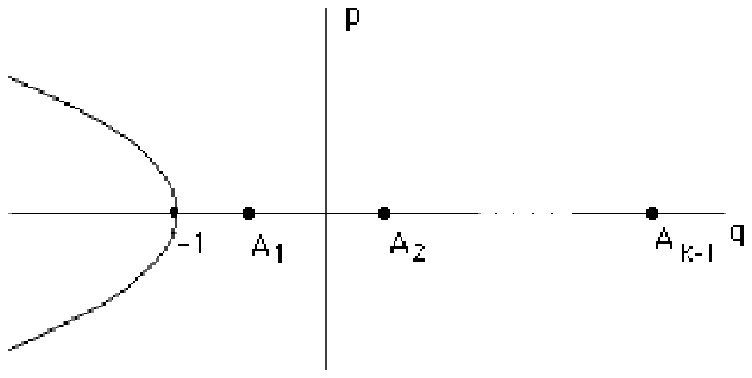}
\begin{center}
Figure~2: \footnotesize{Phase space curve $p(q,E=0)$ for (2,2k-1) minimal string}
\end{center}
\end{figure}

\section{Quantization of Minimal Strings}

As a first step towards the quantization of minimal strings we will consider the 
semi-classical quantization of the auxiliary mechanical model. This approach leads to map 
higher genus corrections to the minimal string theory into quantum corrections of the 
auxiliary mechanical system. More precisely we will identify the wave function of the 
eigenstate $E=0$ with the all genus FZZT partition function.

\begin{figure}
\centering
\includegraphics{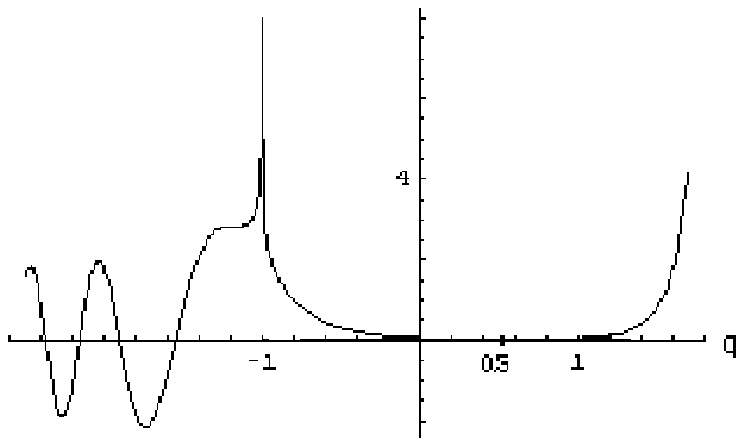}
\begin{center}
Figure~3: \footnotesize{$\psi_{E=0}$ in WKB approximation for the (2,3) system.}
\end{center}
\end{figure}

In order to quantize the analog mechanical model we first define the   associated 
Heisenberg algebra \ben [\hat{q},\hat{p}] = i\hbar \een The meaning of this Heisenberg 
algebra as well as the Planck constant will become clear in the context of matrix models 
in next section. For the given value $E=0$ we define in WKB
\ben Z_{FZZT}(x) \equiv \psi_{E=0}(q)\approx \Big{\arrowvert}   
\frac{\partial^{2}S}{\partial_{q}   \partial_{E}}(q,E=0)\Big{\arrowvert}^{\frac{1}{2}}   
e^{\frac{i}{\hbar}S(q,E=0)} \een
In the simplest $(2,1)$ model we can easily get the exact   eigenstate. It is given by 
\ben \label{airy21} \psi_{E=0}(q)= Ai \left( \frac{q+ 1}{2^{1/3}\hbar^{2/3}} \right) \een 
the Airy function. For the $(2,3)$ model we get in WKB the function   depicted in figure 
3. Notice that this function diverges at $q=   \infty$. Due to reasons we will explain 
below, we take the   asymptotic behavior ansatz in such a way that, for the $(2,2k-1)$   
model, $\psi_{E=0}(q)$ vanishes at $q=+\infty$ for $k$ odd. For $k$   even, which 
corresponds to nonperturbatively inconsistent models,   $\psi_{E=0}(q)$ diverges at 
$q=+\infty$.

\subsection{The Quantum Corrected Target and The Fold Catastrophe}

In order to see the quantum fate of the curve $M_{p,q}$, defined by   $y=y(x)$ or by the 
mechanical analog torus $p=p(q,E)$, we use the   quantum corrections to the Wigner 
distribution function $f(p,q)$   \cite{Berry}. As it is well known, the Wigner function is 
the   quantum mechanical generalization of the classical Boltzmann   distribution function 
on phase space. It is defined by \ben
f_{E}(p,q) = \frac{1}{\pi\hbar} \int dX
e^{-\frac{2i}{\hbar}pX} \psi_{E}(q+X)\psi_{E}^{*}(q-X)
\een
The reason we are interested in the Wigner function
is because it is a very natural way to study quantum deformations of   the curve $p(q,E)$ 
on phase space. In fact in the classical limit it is easy to see that   $f(p,q)$ is a 
delta function on the curve $p(q,E)$.

Using the WKB approximation for $\psi_{E}(q)$ we get for the Wigner   function the 
following integral representation \ben \label{cuatro}
f_{E}(p,q) = \frac{1}{\pi h}\int_{-\infty}^{+\infty} dX   \frac{\exp{\left[ 
\frac{i}{\hbar} \int_{q-X}^{q+X} p(q^\prime ,   E)dq^\prime -2pX 
\right]}}{\Big{\arrowvert} \frac{\partial I}{\partial   p}\left( q+X,p(q+X)  \right)  
\frac{\partial I}{\partial p}\left(  
q-X,p(q-X)  \right)  \Big{\arrowvert}^{\frac{1}{2}}}
\een
Given a generic point $(p,q)$, the saddle point approximation to  
(\ref{cuatro}) is determined by the two solutions $X_{1}$ and $X_{2}$ of   \ben 
\label{cinco}
p(q+X,E) + p(q-X,E) =2p
\een
It is easy to see that these saddle points coalesce when $(p,q)$ is   on the curve 
$p=p(q,E)$. The net effect of this is that the curve   $p=p(q,E)$ defines a fold 
catastrophe for the Wigner function. The   resolution of this catastrophe is given in 
terms of the Airy   function \ben \label{airypeak}
f(p,q) =\frac{\sqrt{2}\left[ \frac{3}{2}A(q,p)  \right]^\frac{1}{6}   Ai\left( -\left[ 
\frac{3A(q,p)}{2 \hbar}\right]^\frac{2}{3}  \right)    }{\pi \hbar^{2/3} \Big{\arrowvert}  
\frac{\partial I}{\partial q}(2)   \frac{\partial I}{\partial p}(1)-\frac{\partial 
I}{\partial  
q}(1)\frac{\partial I}{\partial p}(2)    \Big{\arrowvert}^{1/2}}
\een
where $A(q,p)=\int_{q-X}^{q+X} p(q^\prime , E)dq^\prime -2pX$. The   net result for a 
bounded system, with $p=p(q,E)$ a closed curve, is an   oscillatory behaviour for the 
points inside the curve and   exponentially decreasing behaviour for the points in phase 
space   outside the curve. It is important to keep in mind that the Airy   function 
representing the quantum corrections to the Wigner   distribution function near the curve 
$p=p(q,E)$ is a consequence of   the fold catastrophe on this curve and is generic for any 
one   dimensional integrable mechanical model.

\begin{figure}
\centering
\includegraphics{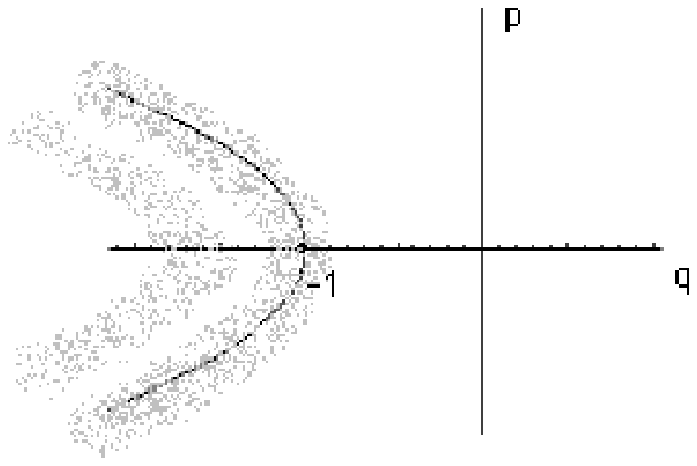}
\begin{center}
Figure~4: \footnotesize{The curve in phase space gets fuzzy by quantum corrections}
\end{center}
\end{figure}

When we translate the curve $p=p(q,E)$ into the associated minimal   string curve 
$y=y(x)$, what we observe is that quantum effects   induce a pattern of fringes on the 
phase space. From this point of   view, where the classical target space of the minimal 
string,   defined by the curve $y=y(x)$, is interpreted as the support of the   Wigner 
function in the classical $h=0$ limit, the effect of quantum   corrections on the target 
space is not to change this curve into   another curve but to spread the curve on the 
phase space in the way   dictated by the type of catastrophe, a fold for one dimensional   
models, defined by the classical curve $y=y(x)$ (see figure 4).

\subsection{Quantum Target and Stokes' Phenomenon}

As we described in the previous section, the resolution of
the fold catastrophe on the curve $p=p(q,E)$ leads to a Wigner   function defined in terms 
of the Airy function. The role of the   Stokes' phenomenon \cite{Maldacena:2004sn}
 in this context is   transparent from a geometrical point of view. From the integral 
representation (\ref{cuatro})   of the Wigner function we observe that, for points in the 
convex   complement of the concave set defined by the curve $p=p(q,E)$, the solutions to 
the saddle point   equation (\ref{cinco}) are complex, leading to an imaginary   
contribution to the exponent in (\ref{cuatro}). The Stokes'   phenomenon implicit in the 
Airy function provides the apropiated   asymptotic form on the convex side, namely 
exponential decay.

For $(2,2k-1)$ models we have seen in section 3 that the
curve $p=p(q,E)$ contains in addition a set of $k-1$ discrete points   on the convex side 
related to the existence of different types of ZZ   branes. In this case the Wigner 
function, as well as the wave   function $\psi_{E}(q)$, has oscillatory behavior in the 
concave   side. Since we can always consider a deformation of the model that   sends the 
discrete points to infinity, obtaining the $(2,1)$ model,   we have exponential decay on 
the convex side infinitesimaly close   to the curve. This implies that the asymptotic form 
on the convex   side at infinity decay exponentially only if $k$ is odd\footnote{The WKB 
expression for the wave function is $\psi_{E=0}\sim (x+1)^{-1/4}\exp{\left[-\hbar^{-1}\int 
ydq \right]}$ with the branch of $y$ in the integral changing 
in every singular point.}.

\section{Matrix Models and The Mechanical Analog}

A non perturbative definition of $(p,q)$ minimal string theory can   be given by the 
double scaling limit (d.s.) of certain matrix models\footnote{See 
\cite{Ginsparg:1993is,Daul:1993bg} and references therein}. In this section we consider 
one matrix models   corresponding to $(2,q=2k-1)$ minimal strings \ben Z_m= 
\frac{1}{\textrm{vol}(U(N))}\int dM e^{-\frac{1}{g_m}\tr V(M)} \een The goal of this 
section is to derive, from the matrix model point   of view, the map defined in 
(\ref{dos}).

\subsection{Wigner Distribution Formalism}

Let us denote $\Pi_{n}(\lambda)= \frac{\lambda^n}{\sqrt{h_n}}+\dots$   the orthonormal 
polynomials of the one matrix model and let us   introduce the complete set of states 
$|\psi_{n}\rangle$ as \ben \langle\lambda|\psi_{n}\rangle= \psi_{n}(\lambda) = 
\Pi_{n}(\lambda) e^{-\frac{V(\lambda)}{2g_{m}}} \een If we define the operator $\hat{q}$ 
by \ben \langle\psi_m|\hat{q}|\psi_n\rangle=\sqrt{\frac{h_m}{h_{m-1}}}\delta 
_{m,n+1}+\sqrt{\frac{h_n}{h_{n-1}}}\delta_ 
{m+1,n}
\een
the recurrence relation of the orthonormal polynomials can be   expressed as \ben
\langle\lambda|\hat{q}|\psi_n\rangle=\lambda\psi_n(\lambda)
\een
that is, $\hat{q}$ acts like the position operator. The   corresponding momentum operator 
$\hat{p}$ can be defined by \ben \langle\lambda |   
\hat{p}|\psi_n\rangle=-ig_m\frac{\partial}{\partial\lambda}\psi_n(\lambda)
\een
One can consider $|\psi_n\rangle$ as defining the Hilbert space of a   one particle 
quantum mechanical system with the coupling constant $g_m$ playing the role of the Planck 
constant $\hbar$. Now, if we   define a mixed state whose density operator is 
\ben\label{rho} \hat{\rho} = \sum_{n=0}^{N-1} \frac{1}{N}|\psi_{n}\rangle   
\langle\psi_{n}| \een we easily get \ben <\frac{1}{N} \tr M^{k}> =   \tr(\hat{\rho} 
\hat{q}^{k}) \een The corresponding Wigner function of this mixed state is \ben
f(p,q) = \frac{1}{\pi \hbar} \int dX e^{-\frac{i p X}{h}}\langle  
q+X|\hat{\rho}|q-X\rangle
\een
from which we obtain that every matrix model correlator can be   obtained from $f(p,q)$ as 
a statistical mean value over the ensemble  
(\ref{rho})
\ben
<\frac{1}{N} \tr M^{k}> = \int dp dq f(p,q) q^{k}
\een
In particular for the resolvent we have
\ben
R(x) = \int d\lambda \frac{\rho_\hbar(\lambda)}{x- \lambda} = \int dpdq f(p,q) 
\frac{1}{x-q} \een which leads to the compact expression \ben
\rho_\hbar(\lambda) = \int dp f(p,\lambda)
\een
for the eigenvalue density.

\subsection{The Mechanical Analog}

Since we are considering this ensemble to correspond to an   equilibrium state, the 
natural thing is to impose the states   $|\psi_m\rangle$ to be stationary states of the 
mechanical system, that is,   eigenstates of a hamiltonian \ben 
\hat{H}|\psi_n>=E_n|\psi_n\rangle \een 
Information about $\hat{H}$ can be derived from the representation of the wave function   
$\psi_N (x)$ as a  matrix model   correlator \ben\label{siete} \frac{1}{\sqrt{h_{N}}} 
e^{-\frac{V(x)}{2g_{m}}}<det(x-M)> =  
\psi_{N}(x)
\een
For $g_m$ small (with 't Hooft coupling $t=g_mN$ fixed)
\ben
<det(x-M)>\approx e^{<tr\log (x-M)>}=e^{N\int^x dx R(x)}
\een
In this regime the resolvent is known to have the structure \ben R(x)\approx 
\frac{1}{2t}\left( V'(x)-y(x) \right) \een from which we obtain \ben 
\frac{1}{\sqrt{h_{N}}} e^{-\frac{V(x)}{2g_{m}}}<det(x-M)> \approx   \frac{1}{\sqrt{h_{N}}} 
e^{-\frac{1}{2g_{m}} \int^x  y(x) dx } \een On the other hand, the first term in WKB approximation for 
$|\psi_N\rangle$ gives \ben
\psi_{N}(x) \approx C e^{-\frac{i}{h} \int^{x}   p(x',E_{N})dx'} \een 
Therefore \ben y(x)= ip(x,E_N)\qquad \int y(x) dx=iS(x,E_N)\qquad 2g_m =\hbar \een which, 
after double-scaling limit, is the matrix model version of   the map (\ref{dos}).

\subsection{The Double Scaling Limit}

Let us briefly recall the definition of double scaling in one matrix   models 
\cite{Ginsparg:1993is}. The relations between the orthogonal polynomials \ben \int 
d\lambda e^{-\frac{V}{g_m}}\Pi^\prime _{n} \lambda \Pi _{n} = n \een \ben \int d\lambda 
e^{-\frac{V}{g_m}}\Pi^\prime _{n} \lambda \Pi _{n} =   \sqrt{r_{n}} \int d\lambda 
e^{-\frac{V}{g_{m}}} \frac{V'}{g_{m}} \Pi   _{n} \Pi _{n-1} \een with $r_n=h_n/h_{n-1}$, 
lead in the large $N \to \infty$ limit to   the equation \ben t\xi = W(r(\xi)) + 
O(\frac{1}{N}) \een where $\xi= \frac{n}{N}$ and   $r(\xi)$ the continuum function 
asoociated with $r_n$. The critical   behavior of the $(2,2k-1)$ minimal string is given 
by adjusting   $V(\lambda)$ in such a way that, at some $r=r_c$, we have \ben W'(r=r_{c})= 
W''(r=r_{c}) = \cdots = W^{k-1}(r=r_{c})= 0 \een If we call $W(r=r_c)=t_c$, the double 
scaling limit is defined by \ben \frac{1}{N}=a^{2+\frac{1}{k}}\kappa
\een
\ben
t_{c} -t\xi = a^{2}\kappa^{-1/(1+\frac{1}{2k})}
\een
\ben
\label{zoom}
\lambda = \lambda_c + a^{\frac{2}{k}}\tilde{\lambda}
\een
\ben
a \to 0
\een
where $\lambda_c$ is the position of the edge of eingenvalue   distribution. In this limit 
the parameter $\kappa$ controls the   genus expansion and becomes the string coupling 
constant $g_s$.

The effect of double scaling over   the mechanical system associated with the matrix model 
is to transform the discrete set of $|\psi_n \rangle$ states into a continuum $|\psi_\xi 
\rangle$ in such   a way that the matrix elements of $\hat{q}$ and $\hat{p}$ have   
continuum indices and become the Lax operator of the system. Due to   the zoom 
(\ref{zoom}) the mechanical system becomes an   unbounded system with quantum corrections 
parametrized by $\kappa$.  The eingenfunction $\psi_{E_N}$, now called $\psi_{E=0}$, which   
is known to give the all genus FZZT partition function, \ben \psi_{E=0}(x)= 
\frac{1}{\sqrt{h_{N}}}   e^{-\frac{V(x)}{2g_{m}}}<det(x-M)> \Big{\arrowvert}_{d.s.} \een 
becomes exactly the Baker-Akhiezer function of the KP hierarchy \cite{Ginsparg:1993is}.

\section{The Fold Catastrophe and Double Scaling}

We have seen in the last section that, from the point of view of the   analog mechanical 
model, the double scaling is equivalent to perform a ``zoom"   near the turning point 
(notice that the turning point of the   mechanical model corresponds to the border of the 
eigenvalue   distribution). Thus, the strategy we will follow to derive the exact 
Baker-Akhiezer function is : \begin{enumerate} \item Starting with the matrix model, we 
define the corresponding   bounded classical mechanical model. \item Near the classical 
torus $p=p(q,E_N)$ we get a fold   catastrophe for the Wigner function. \item By uniform 
approximation we get a resolution of the fold   catastrophe for the Wigner function near 
the classical torus. \item We perform a ``zoom" in $q$ coordinate near the turning point. 
\item We finally derive the Baker-Akhiezer function by integrating the   Wigner function 
\ben \label{ocho} \int dp f_{E=0}(p,q) = |\psi_{E=0}(q)|^{2} \een \end{enumerate}

Let us exemplify this procedure for the gaussian (2,1) matrix model. The large $N$   
resolvent is in this case
\ben
R(x) = \frac{1}{2t}\left(\lambda- \sqrt{\lambda^{2} -4t}\right)
\een
According to the previous discussion the associated
classical mechanical model is defined by
\ben
p(q,E=E_N)= \sqrt{E_N-q^{2}}
\een
with $E_N=4g_m N$. The torus in phase space, defined by $p=p(q,E)$, is a closed curve with 
the origin as its center of symmetry. From the resolution of the fold catastrophe on this 
curve (\ref{airypeak}) we have
\ben
\int_{-\infty}^{+\infty} dp f_{E_N}(p,q)= \frac{2 \sqrt{2}}{\pi \hbar^{2/3}} 
\int_{0}^{+\infty} dp 
\frac{\left[ \frac{3}{2}A(q,p)  \right]^\frac{1}{6}   Ai\left( -\left[ \frac{3A(q,p)}{2 
\hbar}\right]^\frac{2}{3}  \right)    }{ \Big{\arrowvert}  \frac{\partial I}{\partial 
q}(2)   \frac{\partial I}{\partial p}(1)-\frac{\partial I}{\partial  
q}(1)\frac{\partial I}{\partial p}(2)    \Big{\arrowvert}^{1/2}}
\een
Once we perform the zoom into the turning point $q=+\sqrt{4 g_m N}$ we obtain
\ben
dp=\frac{ \Big{\arrowvert} \frac{\partial p}{\partial q }(q,E_N)  \Big{\arrowvert}^{1/2}  
\left[ \frac{3}{2} A(q,0)  \right]^{1/6}  }   { 2 \sqrt{ -\left[ \frac{3}{2} A(q,p)  
\right]^{2/3} + \left[ \frac{3}{2} A(q,0)  \right]^{2/3} }  } d\left[ -\left[ \frac{3}{2} 
A(q,p)  \right]^{2/3}   \right]
\een
\ben
\Big{\arrowvert}\frac{\partial I}{\partial q}(2)   \frac{\partial I}{\partial 
p}(1)-\frac{\partial I}{\partial  
q}(1)\frac{\partial I}{\partial p}(2)\Big{\arrowvert} \to
2 \Big{\arrowvert} \frac{\partial p}{\partial q} (q,E_N) \frac{\partial I}{\partial p} 
\Big{\arrowvert}    
\een
with
\ben
p(q,E_N) \to p(q,E=0) = \sqrt{ - \frac{q+1}{2}}
\een
From (\ref{ocho}) we get:
\ben
|\psi_{E=0}|^2= \frac{\left[  \frac{3}{2} A(q,0)  \right]^{1/3}}{\pi \hbar^{2/3} 
\big{\arrowvert}\frac{\partial I}{\partial p} \big{\arrowvert}} \int_{-\left[  \frac{3}{2} 
A(q,0)  \right]^{2/3}}^{+\infty} dV \frac{Ai \left( V/\hbar^{2/3}  \right)}{\sqrt{V+ 
\left[  \frac{3}{2} A(q,0)  \right]^{2/3}}}
\een
Using now the magic ``projection identity'' \cite{Berry}
\ben
\int_{-y}^{\infty} dx \frac{Ai(x)}{\sqrt{x+y}} = 2^{\frac{2}{3}} \pi  
Ai^{2}\left(\frac{y}{2^{\frac{2}{3}}}\right)
\een
we get the well known result (\ref{airy21}) \cite{Maldacena:2004sn} for the   
Baker-Akhiezer function in the double scaling limit of the gaussian model. This exercise 
is teaching us that the meaning of the double   scaling is the quantum resolution of the 
fold catastrophe on the   target curve at the classical turning point.

Note that this derivation of the Baker-Akhiezer function is general 
and independent of the type of critical point around which we do 
the double scaling limit. In the case of a critical point of order one ($(2,1)$ model) the uniform 
approximation is simple and after the double scaling limit the function 
$\psi_{E=0}$ can be written in terms of the Airy function. On the other hand, 
in the case of $(2,2k-1)$ models the resolution of 
the WKB singularity at the classical turning point is  more complicated due to the fact that the classical turning point is of higher order. However, from the discussion in section 5.2 and 5.3, we conclude that the quantum resolution of the WKB singularity at the classical turning point encodes the same information as the double scaling limit and, therefore, as the Baker-Akhiezer function.

\section{ZZ Brane Backgrounds and Quantization Rules}

We have seen that, for $(2,2k-1)$ models, the curve $p=p(q,E_{N})$ near the turning point 
$q=q_t$ is of the form
\ben
p^{2} \sim (q-q_t)^{2k-1}
\een
The singularity at the turning point is therefore a cusp for pure gravity $(2,3)$ model   
and a ramphoid cusp for the $(2,5)$ model. We can now consider the blow up resolution of 
these singularities. In the case of the cusp we only need to blow up once, which means one   
exceptional divisor. In the case of the ramphoid cusp the singularity can be resolved by 
blowing up twice. In general what we observe is that the resolution of the turning point 
singularity requires $k-1$ blow-ups. As it is well known, in this $(2,2k-1)$ models there 
are ZZ branes of type $(1,l)$ for $l= 1,\ldots,k-1$ \cite{Seiberg:2003nm}. Thus the number 
of different types of ZZ branes is determined by the turning point singularity.

Let us consider backgrounds with ZZ branes. The deformation of the curve $M_{2,2k-1}$ due 
to a background with ZZ branes can be derived at first order from the FZZT-ZZ annulus   
amplitude \cite{Kutasov:2004fg}. This deformation is given by \ben \delta y^{2} = 
-2^{2k-3} \sum_{l=1}^{k-1} g_s n_l \sqrt{x_{1,l}+1}   \prod_{n \neq l} (x-x_{1,n}) \een 
where $n_l$ is the number of ZZ branes of the type $(1,l)$.

From now on we will consider only models with $k$ odd that are well   defined non 
perturbatively. The first thing to observe is that, for the matrix model effective 
potential, defined by \ben
V_{eff}(x) = Re[+\int^{x} y dx]
\een
backgrounds with $n_{l}$ ZZ branes of type $(1,l)$ with l even   correspond to have 
$n_{l}$ eigenvalues at the stable saddle points   i.e the minima of $V_{eff}$, while 
backgrounds with ZZ branes of   type $(1,l)$ with l odd correspond to have some 
eigenvalues at the   maxima of $V_{eff}$ \cite{Sato:2004tz}. This difference between ZZ 
backgrounds   is reflected in the classical mechanical model in a very neat way.   As we 
have already discussed, the classical curve on phase space for   the $(2,2k-1)$ model 
contains a set of $k-1$ points on the $q$ axis.   These points are in correspondence with 
the different types of ZZ   branes. Once we include the deformation of the curve due to 
the   presence of ZZ branes what we get
is:
i) for backgrounds with ZZ branes of type $(1,l)$ with
l even the corresponding point becomes a closed curve (see figure 5)   in phase space,
ii) for
backgrounds with ZZ branes of type $(1,l)$ with l odd the   corresponding point 
disappears. The reason for this phenomenon is quite simple. In fact the deformation   of 
$y^{2}$ due to the presence of ZZ branes is to lift the saddles   $(1,l)$ above or bellow 
the fixed energy line $y^{2}=0$ for l odd   and even respectively (see figure 6).

\subsection{Quantization Rules}

\begin{figure}
\centering
\includegraphics{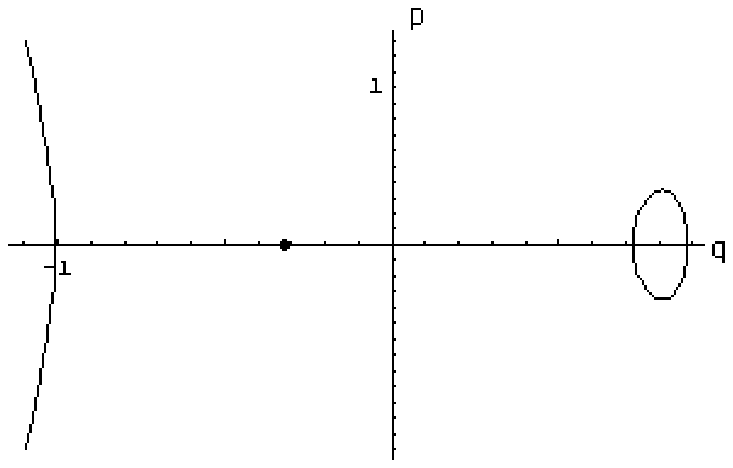}
\begin{center}
Figure~5: \footnotesize{Deformed (2,5) model curve $p(q,E=0)$ in a background with 
$ZZ_{1,2}$ branes.}
\end{center}
\end{figure}

Let us now consider the case of $(1,l)$ ZZ branes with l even. The   corresponding curve 
$p=p(q,E)$ in phase space at the classical limit   contains a closed curve $\Upsilon$ 
around the point $(1,l)$ (see figure 5). This   automatically induces a quantization rule. 
A nice way to derive this   quantization rule is imposing that the Wigner function should 
be   single valued \cite{Berry}. If we consider a closed path passing through the center 
of symmetry of $\Upsilon$ we obtain in saddle   point approximation a phase factor for  
the Wigner function \ben e^{-\frac{i}{h} \oint p dq + \pi i} \een which leads to the 
quantization condition \ben \label{quco}\frac{1}{2\pi} \oint p dq = (n+\frac{1}{2}) \hbar 
\een This quantization rule leads to the identification of the quantum   number $n$ with 
the number of $(1,l)$ ZZ branes with l even, and to the concrete identification between 
the Planck constant $\hbar$ of   the mechanical model and the minimal string coupling 
constant $g_s$.

\begin{figure}
\centering
\includegraphics{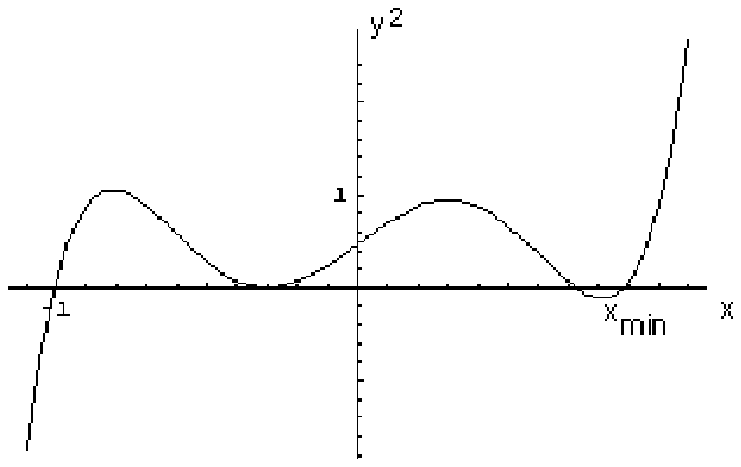}
\begin{center}
Figure~6: \footnotesize{Mechanical system potential after $g_s n_2 = 0.01$ deformation.}
\end{center}
\end{figure}

As an example, we can see these relations in more detail for the $(2,5)$ model. In this 
case the deformed curve is given by (see figure 6)
\ben
y^2=8 \left[  (x+1) (x-x_{1,1})^2(x-x_{1,2})^2-g_s n_2 \sqrt{x_{1,2}+1} (x-x_{1,1})   
\right]
\een
with $x_{1,1}=\frac{1-\sqrt{5}}{4}$ and $x_{1,2}=\frac{1+\sqrt{5}}{4}$. If we consider 
small deformations $g_s n_2 \ll 1$ and expand around the (1,2) ZZ brane minimum we obtain 
the quadratic potential
\ben
y^2 \approx -\omega_1 g_s n_2 + \frac{\omega_2^2}{4}(x-x_{\textrm{min}})^2
\een
with $\omega_1 \simeq \omega_2 \simeq 12.3$. The level $E=0$ satisfy the quantization 
condition (\ref{quco}) only if
\ben
\omega_1 g_s n_2=\hbar \omega_2 n
\een
from which we obtain that $n_2$ is a natural number and
\ben
\frac{\hbar}{g_s}=\frac{\omega_1}{\omega_2} \simeq 1
\een

\section{Summary and Open problems}

In this paper we have suggested a general procedure to approach the quantum gravity 
effects on the target space of minimal strings. This procedure is based on the 
semiclassical mechanics in phase space as described by the Wigner function. The non 
perturbative effects encoded in the exact double scaling limit are in this scheme related 
to the type of singularity of the Wigner function on the classical target. As an example 
we have derived the double scaling limit of the gaussian model. For more general 
$(2,2k-1)$ models we have described the type of singularity at the turning point. It 
remains to work out in full detail the resolution of the corresponding Wigner function in 
these cases. We have also derived a semiclassical quantization rule for ZZ branes, that 
obviously is not taking into account the tunneling unstabilities of these potentials. 
Concrete results on these tunneling rates requires the exact resolution of the Wigner 
function at the turning point for $(2,2k-1)$ models.     

From the point of view of topological strings on a Calabi-Yau defined by $uv+F(x,y)=0$, a 
similar analysis can be done for the system associated to the curve $F(x,y)=0$. The so 
defined Wigner function is very reminiscent of the Witten index for a BPS black hole 
\cite{Ooguri:2004zv}. We hope to address this fascinating issue in a future note. 

\acknowledgments
This work was partially supported by  Plan Nacional de Altas 
Energ\'\i as, Grant FPA2003-02877. 
The work of S.M. is supported by the Ministerio de Educaci\'on y 
Ciencia through FPU Grant 
AP2002-1386. The work of P.R. is supported by a UAM fellowship. 

\bibliographystyle{JHEP-2}

\begin{thebibliography}{77}

\bibitem{Dorn:1994sv}
  H.~Dorn and H.~J.~Otto,
  ``Some conclusions for noncritical string theory drawn from two and three
  point functions in the Liouville sector,''
  arXiv:hep-th/9501019.
\\
  J.~Teschner,
  ``On the Liouville three point function,''
  Phys.\ Lett.\ B {\bf 363} (1995) 65
  [arXiv:hep-th/9507109].
\\
  A.~B.~Zamolodchikov and A.~B.~Zamolodchikov,
  ``Structure constants and conformal bootstrap in Liouville field theory,''
  Nucl.\ Phys.\ B {\bf 477} (1996) 577
  [arXiv:hep-th/9506136].
\\
  V.~Fateev, A.~B.~Zamolodchikov and A.~B.~Zamolodchikov,
  ``Boundary Liouville field theory. I: Boundary state and boundary  two-point
  function,''
  arXiv:hep-th/0001012.
\\
  J.~Teschner,
  ``Liouville theory revisited,''
  Class.\ Quant.\ Grav.\  {\bf 18} (2001) R153
  [arXiv:hep-th/0104158].
\\
  A.~B.~Zamolodchikov and A.~B.~Zamolodchikov,
  ``Liouville field theory on a pseudosphere,''
  arXiv:hep-th/0101152.
\\
  B.~Ponsot and J.~Teschner,
  ``Boundary Liouville field theory: Boundary three point function,''
  Nucl.\ Phys.\ B {\bf 622} (2002) 309
  [arXiv:hep-th/0110244].

  
\bibitem{Seiberg:2003nm}
  N.~Seiberg and D.~Shih,
  ``Branes, rings and matrix models in minimal (super)string theory,''
  JHEP {\bf 0402} (2004) 021
  [arXiv:hep-th/0312170].


\bibitem{McGreevy:2003kb}
  J.~McGreevy and H.~Verlinde,
  ``Strings from tachyons: The c = 1 matrix reloaded,''
  JHEP {\bf 0312} (2003) 054
  [arXiv:hep-th/0304224].
\\
  E.~J.~Martinec,
  ``The annular report on non-critical string theory,''
  arXiv:hep-th/0305148.
\\
  I.~R.~Klebanov, J.~Maldacena and N.~Seiberg,
  ``D-brane decay in two-dimensional string theory,''
  JHEP {\bf 0307} (2003) 045
  [arXiv:hep-th/0305159].
\\
  J.~McGreevy, J.~Teschner and H.~Verlinde,
  ``Classical and quantum D-branes in 2D string theory,''
  JHEP {\bf 0401} (2004) 039
  [arXiv:hep-th/0305194].
\\
  I.~R.~Klebanov, J.~Maldacena and N.~Seiberg,
  ``Unitary and complex matrix models as 1-d type 0 strings,''
  Commun.\ Math.\ Phys.\  {\bf 252} (2004) 275
  [arXiv:hep-th/0309168].
\\
  D.~Gaiotto and L.~Rastelli,
  ``A paradigm of open/closed duality: Liouville D-branes and the  Kontsevich
  model,''
  arXiv:hep-th/0312196.
\\
  M.~Hanada, M.~Hayakawa, N.~Ishibashi, H.~Kawai, T.~Kuroki, Y.~Matsuo and T.~Tada,
  ``Loops versus matrices: The nonperturbative aspects of noncritical string,''
  Prog.\ Theor.\ Phys.\  {\bf 112} (2004) 131
  [arXiv:hep-th/0405076].
\\
  J.~Ambjorn, S.~Arianos, J.~A.~Gesser and S.~Kawamoto,
  ``The geometry of ZZ-branes,''
  Phys.\ Lett.\ B {\bf 599} (2004) 306
  [arXiv:hep-th/0406108].


  
  
\bibitem{Maldacena:2004sn}
  J.~Maldacena, G.~W.~Moore, N.~Seiberg and D.~Shih,
  ``Exact vs. semiclassical target space of the minimal string,''
  JHEP {\bf 0410} (2004) 020
  [arXiv:hep-th/0408039].

\bibitem{Ooguri:2004zv}
  H.~Ooguri, A.~Strominger and C.~Vafa,
  ``Black hole attractors and the topological string,''
  Phys.\ Rev.\ D {\bf 70} (2004) 106007
  [arXiv:hep-th/0405146].
\\
  H.~Ooguri, C.~Vafa and E.~Verlinde,
  ``Hartle-Hawking wave-function for flux compactifications,''
  arXiv:hep-th/0502211.

\bibitem{Berry}
  M.~V.~Berry,
  ``Semi-classical mechanics in phase space: A study of Wigner's function,''
  Phil.\ Trans.\ Roy.\ Soc.\ Lond.\ A {\bf 287} (1977) 237.
    
\bibitem{Ginsparg:1993is}
  P.~H.~Ginsparg and G.~W.~Moore,
  ``Lectures on 2-D gravity and 2-D string theory,''
  arXiv:hep-th/9304011.
\\
  P.~Di Francesco, P.~H.~Ginsparg and J.~Zinn-Justin,
  ``2-D Gravity and random matrices,''
  Phys.\ Rept.\  {\bf 254}, 1 (1995)
  [arXiv:hep-th/9306153].

\bibitem{Daul:1993bg}
  J.~M.~Daul, V.~A.~Kazakov and I.~K.~Kostov,
  ``Rational theories of 2-D gravity from the two matrix model,''
  Nucl.\ Phys.\ B {\bf 409} (1993) 311
  [arXiv:hep-th/9303093].

  
\bibitem{Kutasov:2004fg}
  D.~Kutasov, K.~Okuyama, J.~w.~Park, N.~Seiberg and D.~Shih,
  ``Annulus amplitudes and ZZ branes in minimal string theory,''
  JHEP {\bf 0408} (2004) 026
  [arXiv:hep-th/0406030].

\bibitem{Sato:2004tz}
  A.~Sato and A.~Tsuchiya,
  ``ZZ brane amplitudes from matrix models,''
  JHEP {\bf 0502}, 032 (2005)
  [arXiv:hep-th/0412201].

  
\end{thebibliography}

\end{document}